\documentclass[11pt, fceqn]{article}
 \usepackage{amsfonts}
 \usepackage{flafter}
 \usepackage{amssymb,graphics,graphicx,subfigure,caption2,rotating}
 \usepackage{amsmath, latexsym}
\usepackage[amsmath,thmmarks]{ntheorem}
 \usepackage[top=3.5cm]{geometry}
\begin{document}
\date{}
\title{Separability criteria based on the realignment of density matrices and reduced density matrices}
\author{Shu-Qian Shen$^1$\thanks{E-mail: sqshen@upc.edu.cn.}, Meng-Yuan Wang$^1$, Ming Li$^1$, Shao-Ming Fei$^{2,3}$\\
{\small{\it $^1$College of Science, China University of Petroleum, 266580 Qingdao, P.R. China}}\\
{\small{\it $^2$School of Mathematical Sciences, Capital Normal University, 100048 Beijing, P.R. China}}\\
{\small{\it $^3$Max-Planck-Institute for Mathematics in the Sciences, 04103 Leipzig, Germany}}}
\maketitle
\begin{abstract}
By combining a parameterized Hermitian matrix, the realignment matrix of the bipartite density matrix $\rho$ and the
 vectorization of its reduced density matrices, we present a family of separability criteria, which are stronger than
  the computable cross norm or realignment (CCNR) criterion. With linear contraction methods, the proposed criteria can
  be used to detect the multipartite entangled states that are biseparable under any bipartite partitions. Moreover, we show by examples that the
  presented multipartite separability criteria can be more efficient than the corresponding multipartite
  realignment criterion based on CCNR, multipartite correlation tensor criterion and multipartite covariance matrix criterion.

\end{abstract}

\section{Introduction}
Quantum entanglement, as an intrinsical feature of quantum mechanics, provides the basic physical resource in quantum
 information and computation \cite{Nielsen2010}. It leads to a fundamental problem of how to distinguish
 between entangled states and separable states. Nevertheless this problem is extremely difficult to solve and
 has been proven to be NP-hard \cite{Gurvits2003-1}. In the last decades, a variety of operational methods have been
 proposed to detect entanglement such as the positive partial transpose (PPT) criterion or Peres-Horodecki criterion
 \cite{ppt,Horodecki1996}, realignment criteria \cite{realignment,Chen2003,Zhang2008}, correlation matrix or tensor
 criteria \cite{Vicente2007,correlation,multi-correlation,multi-correlation2}, covariance matrix criteria
 \cite{covariance,covariance-multi}, entanglement witnesses \cite{Horodecki1996,witness}, separability criteria via
 measurements \cite{measurement} and so on; see, e.g., \cite{survey} for comprehensive surveys.

Among the criteria mentioned above, the most popular one is the PPT
criterion \cite{ppt}, which bases on the fact that the partial
transpose of a separable state is positive semidefinite. Moreover,
this criterion is sufficient and necessary for the separability of
$m\times n$ quantum sates with $mn\le 6$ \cite{Horodecki1996}.
However, it only provides necessary conditions for separability of
states with higher dimensions, since there exist entangled $2\times
4$ and $3\times 3$ states with positive partial transposes
\cite{Horodecki1997}. Thus, it is crucial to check whether a given
 PPT state is entangled or not with $mn>6$.

Another well-known one is the computable cross norm or realignment (CCNR) criterion \cite{realignment,Chen2003}, which is
very easy to apply and shows a dramatic ability to detect many PPT entangled states. The multipartite case of this criterion
was considered in \cite{Horodecki2006-2}. The authors showed that the partial realignment can detect the tripartite entangled
states with biseparability under any bipartite partitions, and that the PPT criterion and the CCNR criterion are equivalent to the
permutations of the density matrix's indices. After that, the generalizations of CCNR criterion were investigated in
\cite{generalizations}. In \cite{symmetric}, the authors made use of the symmetric function of Schmidt coefficients to improve the CCNR criterion further.
Recently, the CCNR criterion was used to study the entanglement conditions for any two-mode continuous-variable state
with permutational symmetry \cite{Jiang2014}.

In \cite{Zhang2008}, Zhang et al. presented a separability criterion (for simplicity, we call it the Z-R criterion) based
on the entry realignment of $\rho-\rho^A\otimes \rho^B$, which was shown to be strictly stronger than the CCNR criterion
 and the correlation matrix criterion \cite{Vicente2007}. A generalization of Z-R criterion was studied in \cite{Aniello2008}.
However, the Z-R criterion is still a strong one.

 In this paper, based on a parameterized Hermitian matrix, the realignment matrix of the bipartite density matrix $\rho$ and
 the vectorization of reduced density matrices $\rho^A$ and $\rho^B$, we construct realignment matrices with larger
 scales. Then, separability criteria for bipartite quantum systems that are stronger than the CCNR criterion are presented.
Meanwhile, the new criteria exhibit
  comparative detection abilities of entanglement compared with the Z-R criterion. Finally, by linear contraction methods introduced
  in \cite{Horodecki2006-2}, the proposed criteria are valid to detect the multipartite entanglement states that
  are biseparable under any bipartite partitions, while the Z-R criterion fails to be applied in a similar way. Moreover, two examples show that
  the obtained multipartite separability criteria can outperform the corresponding multipartite realignment criterion based on CCNR \cite{Horodecki2006-2},
  multipartite correlation tensor criterion \cite{multi-correlation,multi-correlation2}, and multipartite covariance matrix criterion \cite{covariance-multi}.

The remainder of the paper is arranged as follows. In Section 2, we
first give the realignment methods, and then introduce the new
separability criteria. Theoretical analysis and an example are employed to show the
efficiency of the presented criteria. In Section 3, the proposed
criteria in Section 2 are extended to the multipartite case.
Meanwhile, two examples are supplemented to show the performance of
the multipartite separability criteria. Finally, some concluding
remarks are made in Section 4.

\section{Separability criteria for bipartite states}

For a matrix $X=(x_{ij})\in \mathbb{C}^{m\times n}$, the vector $vec(X)$  is defined as
\[
vec(X)=(x_{11},\cdots,x_{m1},x_{12},\cdots,x_{m2},\cdots,x_{1n},\cdots,x_{mn})^T,
\]
where $T$ stands for the transpose.
Let $Y$ be an $m\times m$ block matrix with $n\times n$ subblocks $Y_{i,j}$, $i,j=1,\cdots,m$. Then the realignment matrix of $Y$ \cite{Chen2003} is defined as
\[
\mathcal{R}(Y)=\left(vec(Y_{1,1})  \cdots, vec(Y_{m,1}),\cdots, vec(Y_{1,m}), \cdots, vec(Y_{m,m}) \right)^T.
\]

For any quantum state  $\rho$ in $\mathbb{C}^{d_A}\otimes \mathbb{C}^{d_B}$, we define
\begin{equation*}
\mathcal{N}_{\alpha,\ell}^{G}(\rho)=\left( {{\begin{array}{*{20}c}
G& {\alpha \omega_{\ell}(\rho^B)^T}  \\
 {\alpha \omega_{\ell} (\rho^A)}  &\mathcal{R}(\rho) \\
\end{array} }} \right),
\end{equation*}
where $G$ is a given Hermitian matrix, $\alpha$ is an arbitrary real
number, $\ell$ is an arbitrary natural number, $\rho^A$ ($\rho^B$)
is the reduced density matrix of the $A$ ($B$) subsystem, and for
any complex matrix $X$,
 \[
 \omega_{\ell} (X) = \underbrace {\left( {\begin{array}{*{20}{c}}
   {vec(X)} &  \cdots  & {vec(X)}  \\
\end{array}} \right)}_{\ell\;\; columns}.
\]

We denote by $||\cdot ||_{\text{tr}}$, $||\cdot||_2$,
$\text{Tr}(\cdot)$ and $E_{m\times n}$ the trace norm (i.e. the sum
of singular values), the spectral norm (i.e. the maximum singular
value), the trace and the $m\times n$ matrix with all entries being
$1$, respectively. The following theorem establishes the new separability criterion
based on $\mathcal{N}_{\alpha,\ell}^{G}(\rho)$ for bipartite states.
 \\
\\
\textbf{Theorem 2.1.} \emph{Let $G-\alpha^2  E_{\ell\times \ell}$ be positive semidefinite. If the state $\rho$
in $\mathbb{C}^{d_A}\otimes \mathbb{C}^{d_B}$ is separable, then}
\[
{\left\| {\mathcal{N}_{\alpha ,\ell }^{G}(\rho )} \right\|_{\text{tr}}} \le 1+\text{Tr}(G).
 \]
\textbf{Proof.} Since $\rho$ is separable, it can be written as a convex combination of pure product states, i.e.,
\[
\rho=\sum\limits_i p_i\rho_i^A\otimes \rho_i^B,
\]
where $0\le p_i\le 1,\sum\nolimits_i p_i=1,$ $\rho_i^A$ and $\rho_i^B$
are pure states of the $A$ and $B$ subsystems, respectively. One gets
\begin{align}
\label{eq:th211}\left\|\mathcal{N}_{\alpha ,\ell }^{G}(\rho )\right\|_{\text{tr}}=\left\|\sum\limits_i p_i
\mathcal{N}_{\alpha,\ell }^{G}(\rho_i^A\otimes \rho_i^B )\right\|_{\text{tr}}\le \sum\limits_i p_i \left\|
\mathcal{N}_{\alpha ,\ell }^{G}(\rho_i^A\otimes \rho_i^B )\right\|_{\text{tr}}.
\end{align}
Thus, we only need to give the upper bound of \[
\left\|\mathcal{N}_{\alpha,\ell}^{G}(\rho_i^A\otimes \rho_i^B )\right\|_{\text{tr}}={\left\| {\left( {\begin{array}{*{20}{c}}
   G & \alpha \omega_{\ell}(\rho_i^B)^T  \\
   {\alpha \omega_{\ell}(\rho_i^A)} & {vec(\rho_i^A)vec(\rho_i^B)^T}  \\
\end{array}} \right)} \right\|_{\text{tr}}},
\]
where we have used the equality \cite{Chen2003}
\[
\mathcal{R}(\rho_i^A\otimes \rho_i^B)=vec(\rho_i^A)vec(\rho_i^B)^T.
\]
 Since the equality $||vec(|x\rangle\langle x|)||_2=|||x\rangle||^2_2$
holds for any vector $|x\rangle$, there exist unitary matrices $U$ and $V$ such that
\begin{align*}
U\;vec(\rho_i^A)&=\left( {\begin{array}{*{20}{c}}
   {\|vec(\rho_i^A)\|_2} &0&  \cdots  & {0}  \\
\end{array}} \right)^T=\left( {\begin{array}{*{20}{c}}
   {1} &0&  \cdots  & {0}  \\
\end{array}} \right)^T,\\
 V\;vec(\rho_i^B)&=\left( {\begin{array}{*{20}{c}}
   {\|vec(\rho_i^B)\|_2} &0&  \cdots  & {0}  \\
\end{array}} \right)^T=\left( {\begin{array}{*{20}{c}}
   {1} &0&  \cdots  & {0}  \\
\end{array}} \right)^T.
\end{align*}
Furthermore, we have
\begin{align}
\label{eq:thm212}\left( {\begin{array}{*{20}{c}}
   {{I_{\ell}}} & {}  \\
   {} & U  \\
\end{array}} \right)\mathcal{N}_{\alpha,\ell }^{G}(\rho_i^A\otimes\rho_i^B )\left( {\begin{array}{*{20}{c}}
   {{I_{\ell}}} & {}  \\
   {} & {{V^\dag}}  \\
\end{array}} \right) = \left( {\begin{array}{*{20}{c}}
   {G} & {\alpha N}  \\
   {\alpha M} & P  \\
\end{array}} \right): = S^G_{\alpha,\ell},
\end{align}
where $I_{\ell}$ is the ${\ell}\times{\ell}$ identity matrix, and
\[M = {\left( {\begin{array}{*{20}{c}}
   1 & 1 &  \cdots  & 1  \\
   0 & 0 &  \cdots  & 0  \\
    \vdots  &  \vdots  &  \ddots  &  \vdots   \\
   0 & 0 &  \cdots  & 0  \\
\end{array}} \right)_{d_A^2 \times \ell}},~
N = {\left( {\begin{array}{*{20}{c}}
   1 & 0 &  \cdots  & 0  \\
   1 & 0 &  \cdots  & 0  \\
    \vdots  &  \vdots  &  \ddots  &  \vdots   \\
   1 & 0 &  \cdots  & 0  \\
\end{array}} \right)_{\ell \times d_B^2}},~
P = {\left( {\begin{array}{*{20}{c}}
   1 & 0 &  \cdots  & 0  \\
   0 & 0 &  \cdots  & 0  \\
    \vdots  &  \vdots  &  \ddots  &  \vdots   \\
   0 & 0 &  \cdots  & 0  \\
\end{array}} \right)_{d_A^2 \times d_B^2}}.\]
The matrix $S^G_{\alpha,\ell}$ can be repartitioned as
\begin{align}
\label{eq:thm213} S^G_{\alpha,\ell}=\left( {\begin{array}{*{20}{c}}
   {\mathcal{W}_{\alpha,\ell}^G} & {0}  \\
   {0 } & 0  \\
\end{array}} \right),
\end{align}
where
\[
\mathcal{W}_{\alpha,\ell}^G=\left( {\begin{array}{*{20}{c}}
   {G} & {\alpha E_{\ell\times 1}}  \\
   {\alpha E_{1\times \ell} } & 1  \\
\end{array}} \right).
\]
Since the matrix $G-\alpha^2
E_{\ell\times \ell}$ is positive semidefinite, from \cite[Theorem 7.7.6]{Horn1985},  $\mathcal{W}_{\alpha,\ell}^G$ is also positive semidefinite. Due to the fact that
the trace norm of a Hermitian positive semidefinite matrix equals to
its trace, we get, by (\ref{eq:thm212}) and (\ref{eq:thm213}),
\begin{align*}
\label{eq:purestate}
\left\|\mathcal{N}_{\alpha,\ell}^{G}(\rho_i^A\otimes \rho_i^B
)\right\|_{\text{tr}}
=\left\|S^G_{\alpha,\ell}\right\|_{\text{tr}}=\left\|\mathcal{W}_{\alpha,\ell}^G\right\|_{\text{tr}}=1+\text{Tr}(G),
\end{align*}
and then, by (\ref{eq:th211}),
 \[
{\left\| {\mathcal{N}_{\alpha ,\ell }^{G}(\rho )} \right\|_{\text{tr}}} \le 1+\text{Tr}(G).
\]
$\hfill \Box$

The CCNR criterion \cite{realignment, Chen2003} claims that any separable state $\rho$ in
$\mathbb{C}^{d_A}\otimes \mathbb{C}^{d_B}$ satisfies the inequality
\[
||\mathcal{R}(\rho)||_{\text{tr}}\le 1.
\]
It is obvious that  Theorem 2.1 reduces to the CCNR criterion when $\alpha$ is chosen to be $0$.
For the case $\alpha\neq 0$, the following proposition shows that Theorem 2.1 is stronger than the CCNR criterion.
\\
\\
\textbf{Proposition 2.1.}\emph{Theorem 2.1 is stronger than the CCNR criterion when $\alpha \neq 0$.}
\\
\\
\textbf{Proof.} For any state $\rho$ in $\mathbb{C}^{d_A}\otimes \mathbb{C}^{d_B}$, we have, by \cite[Lemma 1]{Vicente2007},
\[
||\mathcal{N}_{\alpha,\ell}^G(\rho)||_{\text{tr}}\ge ||G||_{\text{tr}}+||\mathcal{R}(\rho)||_{\text{tr}}=\text{Tr}(G)+||\mathcal{R}(\rho)||_{\text{tr}}.
\]
Thus, if $||\mathcal{N}_{\alpha,\ell}^G(\rho)||_{\text{tr}}\le 1+\text{Tr}(G)$, one has $||\mathcal{R}(\rho)||_{\text{tr}}\le 1$.
$\hfill \Box$\\

By choosing some special parameterized matrices $G$, we
obtain the following two corollaries for detecting entanglement of
bipartite states.
 \\
 \\
 \textbf{Corollary 2.1.} \emph{If the state $\rho$ in $\mathbb{C}^{d_A}\otimes \mathbb{C}^{d_B}$ is separable, then}
\[
{\left\| {\mathcal{N}_{\alpha ,\ell }^{\ell\alpha^2  I_{\ell}}(\rho )} \right\|_{\text{tr}}} \le 1+\ell^2\alpha^2  .
 \]
\textbf{Proof.} We take $G=\ell \alpha^2  I_{\ell}$. Obviously $G-\alpha^2 E_{\ell\times \ell}$ is positive semidefinite.
Then, from Theorem 2.1, the conclusion holds.
 $\hfill \Box$
 \\
 \\
\textbf{Corollary 2.2.} \emph{If the state $\rho$ in $\mathbb{C}^{d_A}\otimes \mathbb{C}^{d_B}$ is separable, then}
\[
{\left\| {\mathcal{N}_{\alpha ,\ell }^{\alpha^2  E_{\ell\times \ell}}(\rho )} \right\|_{\text{tr}}} \le 1+\ell \alpha^2.
 \]
\textbf{Proof.} From $G=\alpha^2 E_{\ell\times \ell}$, we get  $G-\alpha^2 E_{\ell\times \ell}=0$. Hence the conclusion holds by Theorem 2.1. $\hfill \Box$\\

In the following, we consider the problem of the selection of
$\ell$. By adding a row and a column to $G$, we get an
$(\ell+1)\times(\ell+1)$ Hermitian matrix
 \begin{align*}
  \bar G=\left( {\begin{array}{*{20}{c}}
   {\tau} & {\eta^\dag}  \\
   {\eta} & G  \\
\end{array}} \right),
\end{align*}
 where $\eta\in\mathbb{C}^\ell, \tau\in \mathbb{R}$. By an analogous proof of Proposition 2.1, we can derive a more general result immediately.
 \\
 \\
 \textbf{Proposition 2.2.} \emph{If $\bar G-\alpha^2 E_{(\ell+1)\times (\ell+1)}$ is positive semidefinite,
 then the separability criterion} ${\left\| {\mathcal{N}_{\alpha ,\ell+1 }^{\bar G}(\rho )} \right\|_{\text{tr}}} \le 1+\text{Tr}(\bar G)$
 \emph{is stronger than the separability criterion}
 ${\left\| {\mathcal{N}_{\alpha ,\ell }^{G}(\rho )} \right\|_{\text{tr}}} \le 1+\text{Tr}(G).$\\

From Proposition 2.2, it is obvious that Corollaries 2.1-2.2 can detect more entanglement when $\ell$ gets larger.

The Z-R criterion given in \cite{Zhang2008} is based on the realignment of $\rho-\rho^A\otimes \rho^B$. It states that, for any separable
state $\rho$ in $\mathbb{C}^{d_A}\otimes \mathbb{C}^{d_B}$, the inequality
\[
||\mathcal{R}(\rho-\rho^A\otimes \rho^B)||_{\text{tr}}\le \sqrt{(1-\text{Tr}((\rho^A)^2))(1-\text{Tr}((\rho^B)^2))}
\]
must hold. This criterion is also stronger than the CCNR criterion,
but the exact relation between this criterion and Theorem 2.1  needs
to be established.

The following well-known example shows the power of Corollaries 2.1 and 2.2. Nevertheless, we only report the results from Corollary 2.1,
since  Corollary 2.1 has a close performance to Corollary 2.2 by numerical calculations.
\\
\\
\textbf{Example 2.1.}  The following $3\times 3$ PPT entangled state was introduced in \cite{Bennett1999}:
\[
\rho=\frac{1}{4}\left(I_9-\sum\limits_{k=0}^4 |\xi_k\rangle\langle\xi_k|\right),
\]
where
\begin{align*}
|\xi_0\rangle&=\frac{1}{\sqrt{2}}|0\rangle(|0\rangle-|1\rangle),~|\xi_1\rangle=\frac{1}{\sqrt{2}}(|0\rangle-|1\rangle)|2\rangle,~|\xi_2\rangle
=\frac{1}{\sqrt{2}}|2\rangle(|1\rangle-|2\rangle),\\
|\xi_3\rangle&=\frac{1}{\sqrt{2}}
(|1\rangle-|2\rangle)|0\rangle,~ |\xi_4\rangle=\frac{1}{3}(|0\rangle+|1\rangle+|2\rangle)(|0\rangle+|1\rangle+|2\rangle).
\end{align*}
We consider the mixture of $\rho$ with white noise:
\[
\rho_p=\frac{1-p}{9}I_9+p\rho.
 \]
 The CCNR criterion and the Z-R criterion can detect entanglement of $\rho_p$ for $0.8897\le p\le 1$ and  $0.8822\le p\le 1$, respectively.
 The latter entanglement condition $0.8822\le p\le 1$ can also be obtained by Corollary 2.1 when one of the conditions $\ell\ge 12, \alpha\ge 3.4640$
 and $\ell\ge 1, \alpha\ge 11.6590$ holds.
\\

Although Example 2.1 shows that Corollary 2.1 is not better than
the Z-R criterion, their detection abilities for entanglement are
comparative.  More importantly, Theorem 2.1 can be extended to
multipartite states by linear contraction methods
\cite{Horodecki2006-2} directly, see Section 3.

\section{Separability criteria for multipartite states}
In this section, Theorem 2.1 and its corollaries are applied for multipartite systems. An $n$ partite sate $\rho$ in
$ \mathbb{C}^{d_1}\otimes \cdots\otimes \mathbb{C}^{d_n}$ is separable (or fully separable) \cite{Werner1989} if and only if it can be represented as
\[
\rho=\sum\limits_i p_i\rho_i^1\otimes \cdots\otimes \rho_i^n,
\]
 where $p_i\ge 0, \sum\nolimits_i p_i=1$, and $\rho_i^1,\cdots,\rho_i^n$ are pure states of subsystems.

 Under the condition that the linear map $\mathcal{L}_{(k)}$ acting on the $k$ chosen subsystems is contractive on product
 states $\sigma_{j_1}\otimes\sigma_{j_2}\otimes\cdots\otimes\sigma_{j_k}$, where $1\le j_1<j_2<\cdots<j_k\le n$, Horodecki et al.
 \cite{Horodecki2006-2} presented the following separability criterion: if a state $\rho$ in
 $ \mathbb{C}^{d_1}\otimes \cdots\otimes \mathbb{C}^{d_n}$ is separable, then
\begin{align}
\label{eq:multi} ||\mathcal{L}_{(k)}\otimes \mathcal{I}_{(n-k)}(\rho)||_{\text{tr}}\le 1,
\end{align}
where the map $\mathcal{I}_{(n-k)}$ means that the remaining $n-k$ subsystems are left untouched.

To extend Theorem 2.1 to the multipartite case, we define the following map:
\begin{equation}
\label{contractionmap}\mathcal{M}_{\alpha,\ell}^{G}(\rho)=\frac{1}{1+\text{Tr}(G)}\left( {{\begin{array}{*{20}c}
\text{Tr}(\rho)G& {\alpha  \omega_{\ell}(\rho^B)^T}  \\
 {\alpha \omega_{\ell} (\rho^A)}  &\mathcal{R}(\rho) \\
\end{array} }} \right), \;\;\forall\;\; \rho \;\;\text{in}\;\; \mathbb{C}^{d_A}\otimes \mathbb{C}^{d_B},
\end{equation}
where $G$, $\ell$ and $\alpha$ are defined as in Theorem 2.1.
$\text{Tr}(\rho)$ is only to guarantee the linearity of the map in
the extension. From Theorem 2.1, the map (\ref{contractionmap}) is
contractive on any product state $\sigma^A\otimes \sigma ^B$. Thus,
due to Horodeckis' separability criterion (\ref{eq:multi}) for
multipartite systems, we get the following separability criterion based
on $\mathcal{M}_{\alpha,\ell}^{G}$.
\\
\\
\textbf{Theorem 3.1.} \emph{If the state $\rho$ in $ \mathbb{C}^{d_1}\otimes \cdots\otimes \mathbb{C}^{d_n}$ is separable, then}
\[
\left\|\mathcal{M}_{\alpha,\ell}^{G,(2)}\otimes \mathcal{I}_{(n-2)}(\rho)\right\|_{\text{tr}}\le 1,
\]
\emph{where $\mathcal{M}_{\alpha,\ell}^{G,(2)}$ denotes the map $\mathcal{M}_{\alpha,\ell}^{G}$ acting on any chosen 2 subsystems.}
\\

Under the combination of the CCNR criterion and (\ref{eq:multi}), Horodecki et al. \cite{Horodecki2006-2} showed that, if
the state $\rho$ in $ \mathbb{C}^{d_1}\otimes \cdots\otimes \mathbb{C}^{d_n}$ is separable, then
\[
\left\|\mathcal{R}^{(2)}\otimes \mathcal{I}_{(n-2)}(\rho)\right\|_{\text{tr}}\le 1,
\]
where $\mathcal{R}^{(2)}$ denotes the realignment map $\mathcal{R}$ acting on any chosen 2 subsystems. For simplicity, we
call it the H-R criterion. Surprisingly, this criterion can detect the tripartite entangled state which is biseparable under
any bipartite partitions \cite{Horodecki2006-2}. From Proposition 2.1, it can be found that the H-R criterion should be weaker than
Theorem 3.1.

The Z-R criterion is based on the realignment of $\rho-\rho^A\otimes
\rho^B$, but this realignment is not linear on quantum states.
Hence, the Z-R criterion cannot be extended to the multipartite case
by contraction methods. Nevertheless, the Z-R criterion can be
generalized to an analog of permutation separability criterion for
multipartite systems \cite{Zhang2008}. However, the obtained
multipartite separability criterion is only valid for systems of even number of subsystems.

 The following examples illustrate the efficiency of Theorem 3.1 compared with the H-R criterion \cite{Horodecki2006-2}, the
 multipartite correlation tensor criteria \cite[Theorem 1]{multi-correlation} and \cite[Theorem 4]{multi-correlation2}, and
 the multipartite covariance matrix criterion \cite[Proposition 2]{covariance-multi}.
\\
\\
\textbf{Example 3.1.} Consider the tripartite state \cite{Bennett1999}:
\[
\rho_{ABC}=\frac{1}{8}\left(I_8-\sum\limits_{k=1}^4 |\phi_k\rangle\langle\phi_k|\right),
\]
where
\[
|\phi_1\rangle=|0,1,+\rangle,~|\phi_2\rangle=|1,+,0\rangle,~|\phi_3\rangle=|+,0,1\rangle,~|\phi_4\rangle=|-,-,-\rangle,~ \pm=\frac{1}{\sqrt{2}}(|0\rangle\pm |1\rangle).
\]
It was shown in \cite{Bennett1999} that this state is biseparable under any bipartite partitions $A|BC$, $B|CA$ and $C|AB$, but it is still entangled.

To verify the efficiency of Theorem 3.1, we consider the mixture of
$\rho_{ABC}$ with white noise:
\[
\rho^p_{ABC}=\frac{1-p}{8}I_8 +p\rho_{ABC}.
\]
Clearly, $\rho^p_{ABC}$ is also biseparable under any bipartite partitions $A|BC$, $B|CA$ and $C|AB$.

Let $\mathcal{R}_{(BC)}$ and $\mathcal{M}_{\alpha,\ell}^{G,(BC)} $
be the maps $\mathcal{R}$ and $\mathcal{M}_{\alpha,\ell}^{G}$ acting
on $B$ and $C$ subsystems, respectively. By the H-R criterion, the
entanglement of $\rho^p_{ABC}$ for $0.873529 \le p\le 1 $ can be
detected. Meanwhile, from Theorem 3.1 and Corollary 2.1, we choose
$G=\ell\alpha^2 I_{\ell}$. Table 1 displays the entangled conditions
for different values of $\alpha$ and $\ell$. It is easy to see
that Theorem 3.1 is more efficient than the H-R criterion. Moreover,
the detection ability of Theorem 3.1 becomes slightly stronger when
$\alpha$ and $\ell$ get larger.

 However, the corresponding multipartite correlation tensor criteria \cite{multi-correlation,multi-correlation2} and the multipartite
 covariance matrix criterion \cite[Proposition 2]{covariance-multi} cannot detect any entanglement in $\rho^p_{ABC}$.

\begin{table}[htbp]
\centering \begin{tabular} {c|c|c|c|c}\hline
{$\alpha$}&$\ell=1$ & $\ell=10$ &$\ell=100$ &$\ell=500$
 \\ \hline
 {$1$}& $0.845476\le p\le 1$&$0.831017\le p\le 1$ &$0.828701\le p\le 1$  &$0.828483\le p\le 1$
 \\ \hline
 {$10$}&$0.828701\le p\le 1$ &$0.828455\le p\le 1$ &$0.828430\le p\le 1$ &$0.828428\le p\le 1$
 \\ \hline
 {$100$}& $0.828430\le p\le 1$& $0.828428\le p\le 1$& $0.828428\le p\le 1$& $0.828427\le p\le 1$
 \\ \hline
\end{tabular} \caption{\emph{Entanglement conditions of $\rho^p_{ABC}$ from Theorem 3.1 with $G=\ell\alpha^2 I_{\ell}$.}}
\label{tab1}
\end{table}
\noindent\textbf{Example 3.2.} A perturbation of the GHZ state leads to the
 following tripartite qubit state used in \cite{covariance-multi}:
\[
|\psi'_{GHZ}\rangle=\frac{1}{\chi}(|000\rangle+\epsilon |110\rangle+|111\rangle),
\]
where $\epsilon$ is a given real parameter, and $\chi$ denotes the
normalization. We consider the mixture of this state with white
niose:
\[
\rho_{GHZ'}^p=\frac{1-p}{8}I_8+ p|\psi'_{GHZ}\rangle \langle \psi'_{GHZ}|.
\]
 In the numerical demonstration, the maps $\mathcal{R}_{(BC)}$ and $\mathcal{M}_{\alpha,\ell}^{G,(BC)}$ are used for the H-R criterion
 and Theorem 3.1, respectively. From Corollary 2.1, the parameterized matrix $G$ is simply chosen to be $G=\ell\alpha^2I_\ell$ with
 $\alpha=\ell=10$. It should be noted that, for tripartite systems, the multipartite correlation tensor criterion \cite[Theorem 1]{multi-correlation}
 is equivalent to the criterion \cite[Theorem 4]{multi-correlation2}.

 Table 2 gives the results for different values of $\epsilon$.
It follows that Theorem 3.1 outperforms the H-R criterion, the multipartite correlation tensor criterion, and the multipartite
covariance matrix criterion for different values of $\epsilon$.
\begin{table}[htbp]
\centering \begin{tabular} {c|c|c|c|c}\hline
{$\epsilon$}&H-R & M-T &M-C &Theorem 3.1
 \\ \hline
 {$0$}& $0.3344\le p\le 1$& $0.4118\le p\le 1$& $--$ &$0.3334\le p\le 1$
 \\ \hline
 {$10^{-5}$}& $0.3344\le p\le 1$& $0.4118\le p\le 1$& $p=1$ &$0.3334\le p\le 1$
 \\ \hline
 {$10^{-3}$}& $0.3344\le p\le 1$& $0.4118\le p\le 1$& $0.9981\le p\le 1$ &$0.3334\le p\le 1$
 \\ \hline
 {$10^{-1}$}& $0.3340\le p\le 1$& $0.4118\le p\le 1$&  $0.8341\le p\le 1$&$0.3339\le p\le 1$
 \\ \hline
 {$1$}& $0.3899\le p\le 1$& $0.4256\le p\le 1$&  $0.4286\le p\le 1$&$0.3849\le p\le 1$
 \\ \hline
\end{tabular} \caption{\emph{Entanglement conditions of $\rho_{GHZ'}^p$ with different values of $\epsilon$ from the
H-R criterion $($H-R$)$, the multipartite correlation tensor criterion $($M-T$)$, the multipartite covariance matrix
criterion $($M-C$)$, and Theorem 3.1. The symbol ``$--$" denotes that no entanglement can be detected.
}}
\label{tab2}
\end{table}

\section{Conclusion}

In this paper, by introducing a Hermitian matrix and a real
parameter, we first realigned the density matrix and its reduced
density matrices, and then proposed a family of separability criteria,
which, by a strict proof, are stronger than the well-known CCNR
criterion. In general, the new criteria become more efficient when
the involved parameter $\ell$ gets larger. Second, due to the
special choices of $G$, we gave two simple separability criteria, i.e.,
Corollaries 2.1 and 2.2, which are easy to apply and exhibit, by
examples, comparative abilities of entanglement detection compared
with the Z-R criterion. Finally, by linear contraction methods, the
presented criteria were extended to the multipartite case. Two
examples showed that the presented multipartite separability criteria
can be more efficient than the H-R criterion, the multipartite
correlation tensor criterion, and the multipartite covariance matrix
criterion.

There are still many problems that need to be further addressed. For example, the exact relations between Theorem 2.1 and the Z-R
 criterion should be clarified further. How to choose the parameterized matrix and the parameters $\alpha, \ell$ such that the
  proposed criteria can detect more entanglement should be further investigated. Whether some other criteria can be generalized
  and improved by the methods used in this paper is also an interesting problem.

\section*{\bf Acknowledgments}
This work is supported by the NSFC (11105226, 11275131), the
Fundamental Research Funds for the Central Universities (15CX08011A, 24720122013), and the Project-sponsored by SRF for
ROCS, SEM.

\end{document}